\documentclass[12pt,crc,a4paper]{cpa2018}
\usepackage{times}
\usepackage{graphicx}

\usepackage{listings}
\normalfont
\begin{document}
\begin{frontmatter}                           

\setcounter{page}{1}

\title{SpiNNaker 2: A 10 Million Core Processor System for Brain Simulation and Machine Learning%
\thanks{Use template.tex file as a template.}}
\runningtitle{Style Guidelines for CPA Proceedings}

\author[A]{\fnms{Christian} \snm{Mayr}%
\thanks{Corresponding Author: {\em Christian Mayr, TU Dresden, Mommsenstr. 12, 01062 Dresden, Germany}.
Tel.: +49 351 463 42392; Fax: +49 351 463 37794; E-mail: {\tt christian.mayr@tu-dresden.de}.}},
\author[A]{\fnms{Sebastian} \snm{Höppner}}
and
\author[B]{\fnms{Steve} \snm{Furber}}
\runningauthor{B.P. Manager et al.}
\address[A]{Chair of Highly-Parallel VLSI-Systems and Neuromorphic Circuits, Institute of Circuits and Systems, Technische Universität Dresden, Dresden, Germany}
\address[B]{Advanced Processor Technologies Group, School of Computer Science, University of Manchester, Manchester, United Kingdom}

\begin{abstract}
SpiNNaker is an ARM-based processor platform optimized for the simulation of spiking neural networks. This brief describes the roadmap in going from the current SPINNaker1 system, a 1 Million core machine in 130nm CMOS, to SpiNNaker2, a 10 Million core machine in 22nm FDSOI. Apart from pure scaling, we will take advantage of specific technology features, such as runtime adaptive body biasing, to deliver cutting-edge power consumption. Power management of the cores allows a wide range of workload adaptivity, i.e. processor power scales with the complexity and activity of the spiking network. Additional numerical accelerators will enhance the utility of SpiNNaker2 for simulation of spiking neural networks as well as for executing conventional deep neural networks. These measures should increase the simulation capacity of the machine by a factor $>$50. The interplay between the two domains, i.e. spiking and rate based, will provide an interesting field for algorithm exploration on SpiNNaker2. Apart from the platforms' traditional usage as a neuroscience exploration tool, the extended functionality opens up new application areas such as automotive AI, tactile internet, industry 4.0 and biomedical processing. 
\end{abstract}

\end{frontmatter}

\section*{Introduction}
The "Spiking Neural Network Architecture" SpiNNaker is a processor platform optimized for the simulation of neural networks. A large number of ARM cores is integrated in a system architecture optimized for communication and memory access. Specifically, to take advantage of the asynchronous, naturally parallel and independent subcomputations of biological neurons, each core simulates neurons independently and communicates via a lightweight, spike-optimized asynchronous communication protocol~\cite{Furber2013}. Neurons are simulated for a certain timestep (typically 1ms), and then activity patterns exchanged between cores, on the assumption that ‘time models itself’, i.e. an exchange of activity every millisecond is assumed to represent biological real time. This allows the energy efficient simulation of neural network models in real time, with SpiNNaker significantly outperforming conventional high performance computing wrt both these aspects. The first generation of SpiNNaker has been designed by the University of Manchester and is currently operational at its intended maximum system size, i.e. 1 Million ARM processors, as well as in the form of smaller boards in mobile applications (see lower row of images in Fig. 1). One Million cores allow the simulation of spiking neural networks on the order of 1\% of the human brain. Since 2013, Technische Universität Dresden and the University of Manchester have been jointly developing the next generation SpiNNaker2 system in the framework of the EU flagship Human Brain Project (upper row of images in Fig1). Specific target is a scaling of number of cores by a factor of 10, while staying in the same power budget (i.e. 10x better power efficiency). Overall increase in system capacity (simulated numbers of neurons) is expected to be a factor of approx. 50. 

\section{Spinnaker2 capabilities and main new building blocks}

\begin{figure}[hbt]
\includegraphics[width=1\textwidth]{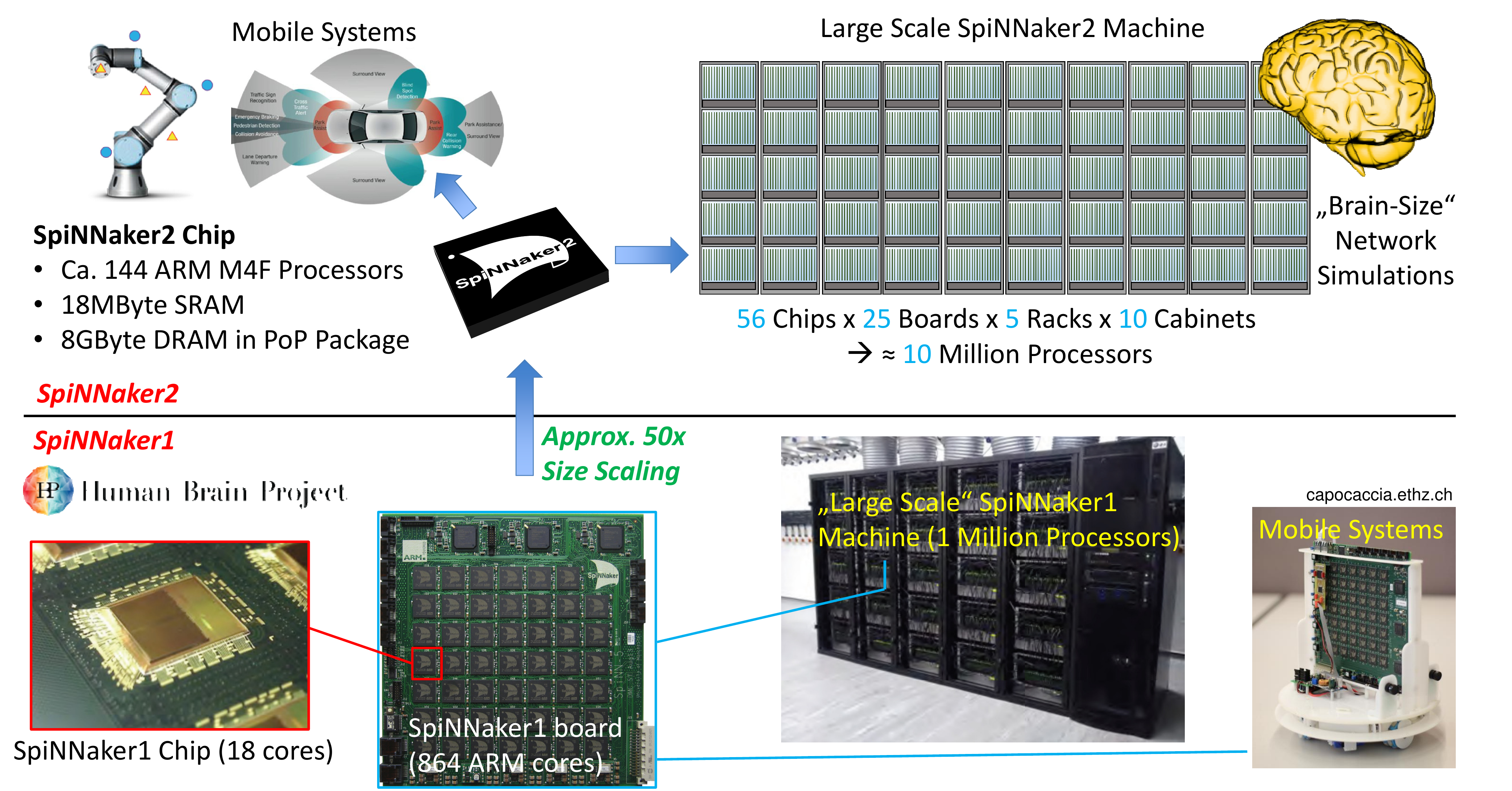}
\caption{First and second generation of the SpiNNaker system. Please note: The above numbers for SpiNNaker2 represent the current state of design and are subject to change.}\label{fig1}
\end{figure}

The 50x capacity scaling is exemplified in the middle of Fig1, i.e. the simulation capacity of a current 48 processor board in SpiNNaker1 is expected to be carried inside a single SpiNNaker2 chip. The additional scaling of 5 above the factor 10 (i.e. increase in core number) is primarily expected from a higher clock rate and numerical accelerators for common synaptic operations. The general approach for SpiNNaker2 can be broken down in four main aspects: 
\textbf{(1) Keep the processor-based flexibility of SpiNNaker1}, which has been a major advantage compared to more streamlined neuromorphic approaches such as Truenorth \cite{Merolla2014} or analog neuromorphic circuits \cite{Petrovici2017,konig2002dedicated}. \textbf{(2) Don’t do everything in software in the processors}, i.e. incorporate numerical accelerators for the most common operations. Current prototypes contain accelerators for exponential functions \cite{Partzsch2017} and derive random numbers from the thermal noise of the clock generators inherent in each core, i.e. at virtually zero circuit overhead \cite{Neumaerker2016}. For an example of the overall benefit of the accelerators in terms of clock cycles, see \cite{Liu2018} and other upcoming papers. Other accelerators, i.e. a log function, are in discussion. \textbf{(3) Use the latest technologies and features for energy efficiency.} We are targeting a 22nm FDSOI technology for SpiNNaker2, with on-chip adaptive body biasing (ABB) enabling reliable operation down to 0.4V under virtually all operating conditions. ABB can narrow down the transistor threshold voltage spread at runtime (i.e. over aging, temperature, manufacturing corner) to enable robust near-threshold logic operation. \textbf{(4) Allow workload adaptivity on all levels.} Similar to the brain, power consumption should be proportional to the task being carried out. Specifically, all processors in SpiNNaker2 operate under dynamic voltage and frequency scaling \cite{Hoeppner2017}. Their operating voltage and frequency are individually adjusted to the load of incoming spikes per every millisecond and the expected clock cycles required to compute this load. Thus, at low computational load times, a task is stretched out over time and can run at less supply voltage, reducing both dynamic and leakage power. Besides this computational adaptivity, communication has also been streamlined. For example, the chip-to-chip links feature very fast power up and down functionality, allowing energy proportionality with regard to the bits transmitted. In addition to the above four main development lines, multiply accumulate arrays (MAC) have been incorporated in the latest prototype chip to enhance the usefulness of SpiNNaker2 beyond simulation of spiking neural networks. With the MACs, the synaptic weighing and accumulation of entire layers of conventional deep neural networks can be offloaded from the processors, freeing them to carry out spike based simulations in parallel. For a discussion on how to parallelize these new network types across processors, see \cite{Liu2018}.

\section{Brain simulation and other applications}

Sheer scaling of SpiNNaker2 compared to SpiNNaker1 should allow the simulation of significantly larger and more complex spiking neural networks. In an upcoming publication, we are showing an implementation of synaptic sampling \cite{Kappel2018}, i.e. the most complex synaptic plasticity ever implemented on SpiNNaker. This implementation is significantly aided by the addition of the new numerical accelerators. In addition, there is scope for extending these spiking simulations to multiscale networks, by using the MACs to simulate entire networks as mesoscopic, ‘black box’ modules. SpiNNaker2 should be able to run models like BioSpaun \cite{Eliasmith2016} on levels of abstraction from multicompartment neurons via spiking point neurons, rate-based neurons up to mesoscopic models. Beyond these extensions of traditional use cases, the upper left corner of Fig. 1 shows other future uses. Specifically, SpiNNaker2 combines high throughput machine learning, sensor-actuator processing with inherent millisecond latency and IoT-device-level energy efficiency, which represents a breakthrough in the field of real-time, mobile human-machine interaction. Targeted applications for SpiNNaker2 in this area include the tactile internet (e.g. tele-learning, robotics interaction), autonomous driving, industry 4.0 (e.g. real time predictive maintenance) or biomedical (e.g. closed-loop neural implants with spiking and machine learning functionality). For an overview of the current SpiNNaker2 development and technical details (MACs, FDSOI approach, accelerators, etc), see \cite{Hoeppner2018}. A last word about the timetable: After the current phase of prototyping, we expect the final SpiNNaker2 chip to become available sometime late 2020/early 2021, so look out for the 2021 Capo Caccia and Telluride neuromorphic workshops. 

\section*{Acknowledgements}

The authors thank ARM  for IP contributions. Discussions at the yearly Capo Caccia workshop have contributed to the prototype designs and benchmarks. The research leading to these results has received funding from the European Union Seventh Framework Programme (FP7) under grant agreement No 604102 and the EU's Horizon 2020 research and innovation programme under grant agreements No 720270 and 785907 (Human Brain Project, HBP).

\bibliographystyle{unsrt}
{\small\bibliography{biblio}}

\end{document}